\documentclass[USenglish,twocolumn]{article}

\usepackage[utf8]{inputenc}
\usepackage[big]{dgruyter}
\usepackage{authblk}
 \usepackage{hyperref}
 \hypersetup{
 	unicode=false,
 	pdftoolbar=true,
 	pdfmenubar=true,
 	pdffitwindow=false,
 	pdfstartview={FitH},
 	pdftitle={My title},
 	pdfauthor={Author},
 	pdfsubject={Subject},
 	pdfcreator={Creator},
 	pdfproducer={Producer},
 	pdfkeywords={keyword1} {key2} {key3},
 	pdfnewwindow=true,
 	colorlinks=true,
 	linkcolor=red,
 	citecolor=blue,
 	filecolor=magenta,
 	urlcolor=cyan
 }

\usepackage{authblk}
\usepackage{pdflscape}
\usepackage{longtable,threeparttable}
\usepackage{rotating} 
\usepackage{colortbl}
\usepackage{multicol}
\usepackage{wasysym}
\usepackage{tablefootnote}
\newcommand{\Msun}{\mbox{$\mathrm{M}_{\rm \odot}$}}
\begin{document}
 
\articletype{Research Article{\hfill}Open Access}

\author*[1,2]{V. Schaffenroth}
\affil[2]{Institute for Astro- and Particle Physics, University of Innsbruck, Technikerstr. 25/8, 6020 Innsbruck, Austria}
\affil[1]{Institute for Astronomy and Astrophysics, Kepler Center for Astro and Particle Physics, Eberhard Karls University, Sand 1,
	72076 Tübingen, Germany E-mail: schaffenroth@astro.uni-tuebingen.de}
\affil[2]{Institute for Astro- and Particle Physics, University of Innsbruck, Technikerstr. 25/8, 6020 Innsbruck, Austria}

\author[3]{B. Barlow}
\affil[3]{Department of Physics, High Point University, 833  Montlieu Ave, High Point, NC 27268  USA}

\author[1]{S. Geier}

\author[4]{M. Vu{\v c}kovi{\'c}}
\affil[4]{Instituto de F\'{i}sica y Astronom\'{i}a, Facultad de Ciencias, Universidad de Valpara\'{i}so, Gran Breta\~{n}a 1111, Playa Ancha, Valpara\'{i}so 2360102, Chile}

\author[5]{D. Kilkenny}
\affil[5]{Department of Physics, University of the Western Cape, Private Bag X17, Bellville 7535, South Africa}

\author[6]{J. Schaffenroth}
\affil[6]{Dr.\,Remeis-Observatory \& ECAP, Astronomical Institute, Friedrich-Alexander University Erlangen-N\"urnberg, Sternwartstr.~7, 96049
Bamberg, Germany}

\title{\huge News from the EREBOS project}

  \runningtitle{News from the EREBOS project}


  \begin{abstract}
{Planets and brown dwarfs in close orbits will interact with their host stars, as soon as the stars evolve to
	become red giants. However, the outcome of those interactions is still unclear. Recently, several brown
	dwarfs have been discovered orbiting hot subdwarf stars at very short orbital periods of 0.065 - 0.096 d.
	More than 8\% of the close hot subdwarf binaries might have sub-stellar companions. This shows that such companions
	can significantly affect late stellar evolution and that sdB binaries are ideal objects to study this influence.
	Thirty-eight new eclipsing sdB binary systems with cool low-mass companions and periods from 0.05 to
	0.5 d were discovered based on their light curves by the OGLE project. In the recently published catalog of eclipsing binaries in the Galactic bulge, we discovered 75 more systems. We want to use this unique and
	homogeneously selected sample to derive the mass distribution of the companions, constrain the fraction
	of sub-stellar companions and determine the minimum mass needed to strip off the red-giant envelope. We
	are especially interested in testing models that predict hot Jupiter planets as possible companions.
	Therefore, we started the EREBOS (Eclipsing Reflection Effect Binaries from the OGLE Survey) project, which
	aims at analyzing those new HW Vir systems based on a spectroscopic and photometric follow up. 
	For this we were granted an ESO Large Program for ESO-VLT/FORS2. Here we give an update on the 
	the current status of the project and present some preliminary results.}
\end{abstract}

  \journalname{Open Astronomy}
\DOI{DOI}
  \startpage{1}
  \received{30 September 2017}
  \revised{18 October 2017}
  \accepted{7 November 2017}

  \journalyear{2017}
  \journalissue{Special Issue on the 8th Meeting on Hot Subdwarf Stars and Related Objects}

\maketitle
\section{Introduction}

Sub-luminous B stars (sdBs) are core helium-burning stars with very thin hydrogen
envelopes and masses around $0.5$ \Msun\, \citep{heber:2009,heber:2016}. To form such an object, the hydrogen
envelope of the red-giant progenitor must be stripped off almost entirely. Since a high fraction of the sdB
stars are members of short-period binaries \citep{maxted:2001}, common envelope ejection
triggered by a close stellar companion is regarded as the most probable formation channel. However, growing evidence suggests
that sub-stellar companions may have a significant influence on the still unclear formation of sdBs as well. 

It has been proposed that planets and brown dwarfs could be responsible for the loss of envelope mass in the
red-giant phase of the sdB progenitors \citep{soker}. As soon as the host star evolves to become
a red giant, close sub-stellar companions must enter a common envelope. Whether those objects are able to
eject the envelope and survive, evaporate or merge with the stellar core depends mostly on their masses. While
planets below $1-10\,{\rm M}_{\rm Jup}$ might not survive the interactions depending on the initial separation, planets and brown dwarfs exceeding this mass might
be able to eject the envelope and survive as close companions. \citet{nt} did a similar calculation for low-mass white dwarfs, coming to the conclusion that in this case companions with masses greater than $\sim 20\, M_{\rm Jup}$ can survive.

Sub-stellar companions around pulsating and close binary sdBs
might have been found \citep[e.g.,][]{silvotti:2007}, but these would be too far away from their
hosts for such interactions. \citet{charpinet:nature} and \citet{silvotti:2014}
discovered possible earth-size objects in close orbits around pulsating sdBs, which might be the remnants of
more massive planets destroyed during the common envelope phase.

The best evidence for interactions with sub-stellar companions is provided by the discovery of three close,
eclipsing sdB binaries (HW Vir systems) with brown dwarf companions. The most successful way to detect
such systems, which also show a characteristic sinusoidal light curve variation caused by the reflection effect, is
by inspecting their light curves \citep[e.g.,][]{vs,barlow:2012}. 

Photometric and spectroscopic follow-up observations of the sdB binary J162256+473051 revealed that the
system is eclipsing with a period of 0.069 d and the companion is likely a brown dwarf \citep{vs:2014_I}. The short period system J082053+000843 (0.096 d) is also eclipsing and the companion has a
mass between 45 and 67 ${\rm M}_{\rm Jup}$ \citep{geier}. Another pulsating HW Vir system (V2008-1753,
0.065 d) with a brown dwarf companion has been discovered most recently \citep{vs:2015a}. Moreover, two brown dwarf candidates around sdB stars have been detected \citep[with
periods $ \sim 0.3\rm\,d$,][]{vs:2014_II}. Since those systems are not eclipsing only minimum
masses can be derived for their companions (0.048 and 0.027 \Msun), which are well below the hydrogen burning
limit.

A photometric follow-up of spectroscopically selected targets (Schaffenroth et al., submitted)
also allowed us to determine the fraction of sub-stellar companions around sdB stars. We derived a fraction of
> 8\% for sub-stellar companions in close sdB binaries. Moreover, they seem to be at least as frequent
as low-mass stellar companions.

This shows that close sub-stellar companions are able to eject a red-giant envelope and that they are much
more frequent than predicted by standard binary population synthesis theory. Due to this high fraction of close sub-stellar
companions hot subdwarfs are very well suited for studying the interactions between stars and brown
dwarfs or giant planets. To understand both the common envelope phase under extreme conditions and the role 
of close-in planets and brown dwarfs for late stellar evolution, we need to study
a homogeneously selected large sample of eclipsing sdB binaries.

\section{The current sample of close sdB binaries with cool companions}

The known sample of eclipsing and non-eclipsing systems is quite inhomogeneous. Most of them have been found in photometric surveys due to their characteristic lightcurve variations \citep[e.g.,][]{vs} or in surveys used to find sdB pulsators \citep[e.g.,][]{jeffrey:2014}. 

\begin{figure}
	\label{logp-mcomp}
	\includegraphics[width=\linewidth]{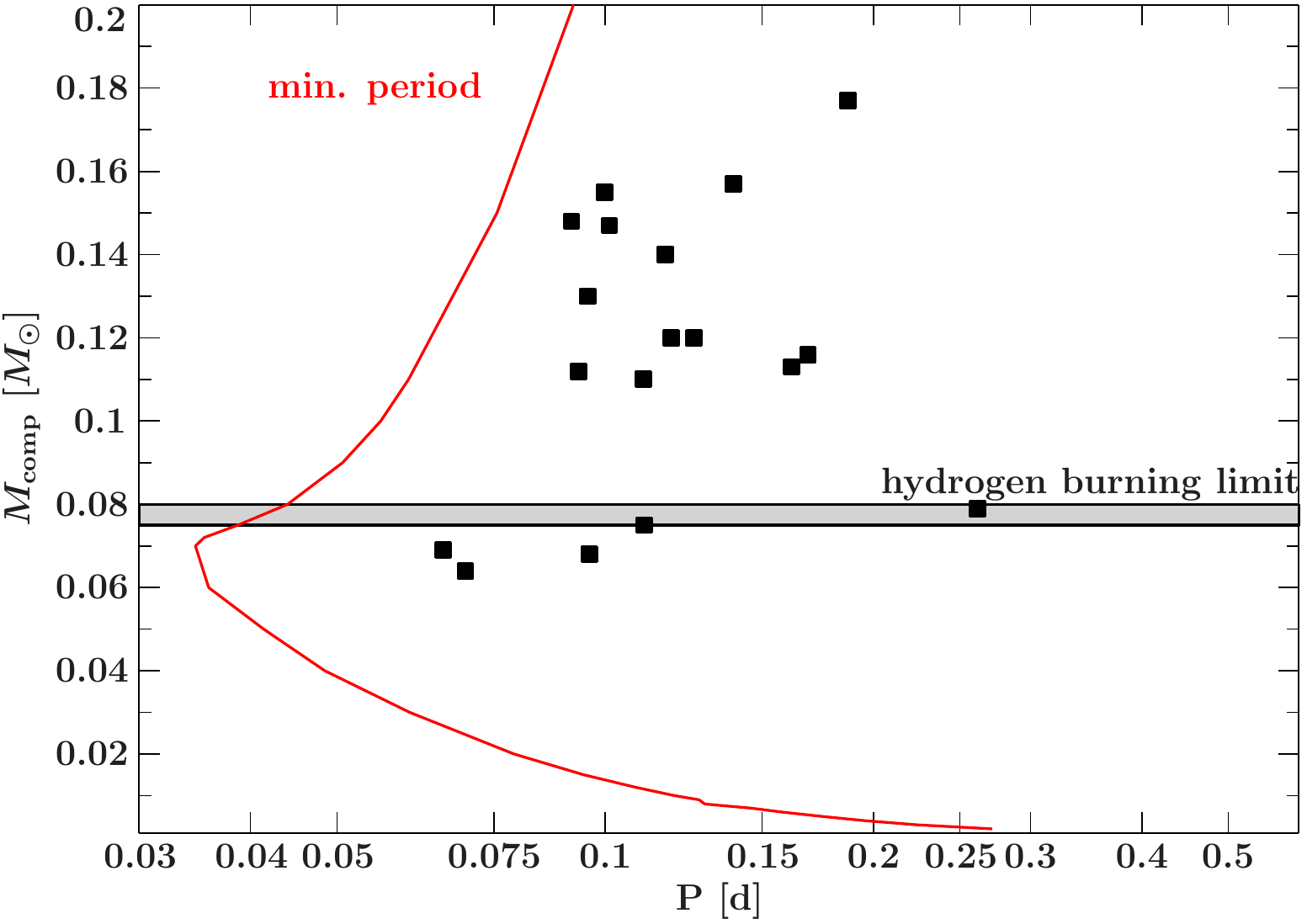}
	\caption{Period-companion mass diagram of the known HW Vir systems. The gray area marks the hydrogen-burning limit. Companions with smaller masses are sub-stellar. The red, solid line gives the minimum period at which a companion of a certain mass can exist. This was calculated assuming that the companions cannot exceed its Roche radius.}
\end{figure}

Figure 1 shows the companion masses of the analyses HW Vir systems plotted against their orbital
periods. There is no obvious correlation between companion mass and period. However, the confirmed sub-stellar
companions seem to be found preferentially in the shortest-period systems. If we assume that the radius of the companion cannot exceed its Roche radius, we can derive the minimum period possible for a system consisting of an sdB and a companion of a certain mass. To derive the Roche radius we used the formula by \citet{eggleton}, which depends on the mass ratio and separation of the binary. For the radius of the companion we used the mass-radius relation by \citet{baraffe:98}. It can be seen that the minimum period for an sdB with a cool, low-mass companion is reached for companions in the brown dwarf mass range. Companions with masses of a few Jupiter masses on the other hand can only exist at longer periods of 0.2 to 0.25 d.

\begin{figure}
	\label{k-m2min}
	\includegraphics[angle=-90,width=\linewidth]{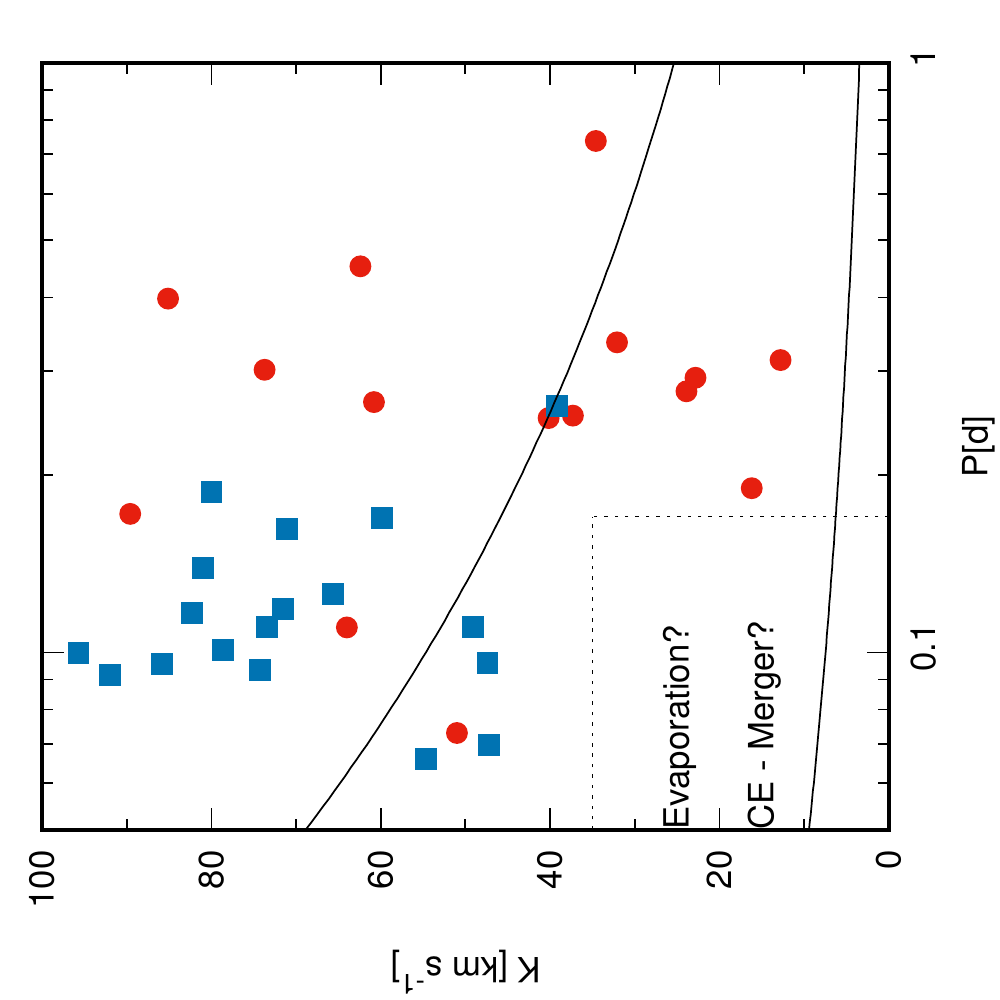}
	\caption{This figure is an updated version of Fig. 4 from \citet{vs:2014_II}. It shows the RV semiamplitudes of all known sdB binaries with reflection
		effects and spectroscopic solutions plotted against their orbital periods. Blue squares mark eclipsing sdB binaries
		of HW Vir type where the companion mass is well constrained. Red circles mark
		systems without eclipses, where only lower limits can be derived for
		the companion masses. The lines mark the regions to the right
		where the minimum companion masses derived from the binary mass
		function (assuming 0.47 \Msun for the sdBs) exceed 0.01 \Msun (lower curve)
		and 0.08 \Msun (upper curve).}
\end{figure}

For non-eclipsing systems, the absolute mass of the companion cannot be determined. Assuming the canonical mass of 0.47~\Msun, however, a minimum mass can be derived. An overview of the 31 sdB binaries
with reflection effects and known orbital parameters is shown in Fig. 2. Although only minimum masses can
be derived for most of the companions, we can use this sample to do some statistics. While most companions
are late M-dwarfs with masses close to $\sim$ 0.1 \Msun, there is no sharp drop below the hydrogen-burning limit. The
fraction of close sub-stellar companions is substantial. An obvious feature in Fig. 2 is the lack of binaries with
periods shorter than $\sim 0.2$ d and $K < 50\,\rm kms^{-1}$ corresponding to companion masses of less than $\sim 0.06\, {\rm M}_{\rm Jup}$.
Since the eclipses of a giant planet and a star close to the hydrogen-burning limit are of equal depth, this
feature is not caused by selection effects. An explanation for the lack of objects at short periods could be that
the companion triggered the ejection of the envelope, but was destroyed during the common-envelope phase. This could also explain the formation of some of the single sdB stars.

\section{The EREBOS project}

To study the properties of the population of hot subdwarf stars with low-mass stellar or sub-stellar companions, a large and homogeneously selected sample of eclipsing binaries is essential. 

Thirty-six new HW\,Vir candidates have been discovered by the OGLE project \citep{ogle,ogle_II}, tripling the number of such objects and providing the first such sample of eclipsing sdB stars. The stars have been identified by their blue colors and their characteristic light curve shapes in the I-band. These light curve shapes and the derived orbital periods leave no doubt that they are indeed HW\,Vir systems. 

Most recently \citep{ogle_III} published a catalog of 450\,000 eclipsing systems in the Galactic bulge. We selected binaries with orbital periods $P<1$ d and applied a color cut V-I $<$ 1 mag. The  resulting lightcurves ($\sim10\,000$) were phased and inspected by eye. In this way we discovered 55 new HW Vir system candidates. 

This sample together with the HW Vir candidates found by \citet{ogle} and \citet{ogle_II} was then used to train a machine-learning algorithm. In this way another 20 systems were identified increasing the sample to 111 HW Vir candidates. This increases the sample of known HW Vir systems by a factor of more than five and is providing a large homogeneously selected sample for the first time. There might be a hidden population of hot white dwarf primaries with cool, low mass companions, especially at longer periods, which have similar lightcurves showing large reflection effects. Therefore our sample is still preliminary at this point.

To investigate this unique sample we started the EREBOS (Eclipsing Reflection Effect Binaries from the OGLE Survey) project that is aimed at measuring orbital and atmospheric parameters for as many of these systems as possible. The most crucial part of the project is time-resolved spectroscopy of these rather faint and short-period binaries.

We were awarded an ESO Large Program of 110h duration with FORS2@VLT to obtain accurate radial velocity curves for the 23 systems with the shortest periods. A combined analysis of the time-resolved spectroscopy together with the lightcurves will allow us to determine the masses of both components and the separations of the systems. 

In addition, we started a photometric follow-up campaign using mostly 2-4m telescopes (ESO-NTT/Ultracam, SOAR/Goodman and SOAR SOI, SAAO-1.9m/SHOC). Moreover, radial velocity curves of five longer-period systems were already taken.



\section{First results}

In Fig. 3 the magnitude distribution of the OGLE HW Vir candidates is shown in comparison to the known HW Vir systems. Most of the OGLE targets have magnitudes of $\sim$ 19 mag, which is much fainter than the known HW Vir systems. 

\begin{figure}
	\label{mag}
	\includegraphics[width=\linewidth]{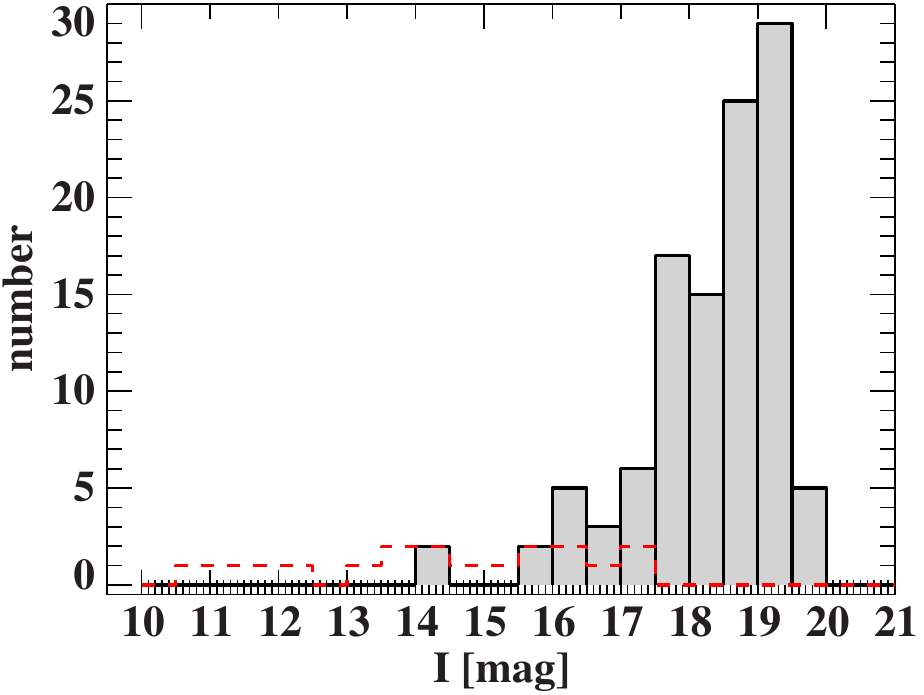}
	\caption{Magnitude distribution of the OGLE HW Vir candidates in comparison to the known HW Vir systems. The OGLE systems are shown in gray and the known systems by the red dashed line.}
\end{figure}

The OGLE sample is still contaminated by hot white dwarf binaries with cool, main sequence companions, which are showing similar lightcurves with large reflection effects. These are post-AGB equivalents to the post-RGB hot subdwarfs. Particularly, targets at longer periods with a reflection effect of large amplitude might not be explainable by hot subdwarf binaries, as they cannot heat up the companion to the temperatures necessary. Hence, the statistical interpretation of our sample is very preliminary and our results have to be treated cautiously, in particular at the longest periods. To solve this issue we are about to get spectroscopic confirmation of all OGLE targets.

Fig. 4 shows the period distribution of the OGLE targets in comparison to the known HW Vir systems, which show a peak around 0.1 d. 
The period distribution of the OGLE targets looks quite different. The maximum is still found around 0.1 d. However, we also have a large population at longer periods from 0.25 to 0.5 d, which was not known before.

\begin{figure}
	\label{period}
	\includegraphics[width=\linewidth]{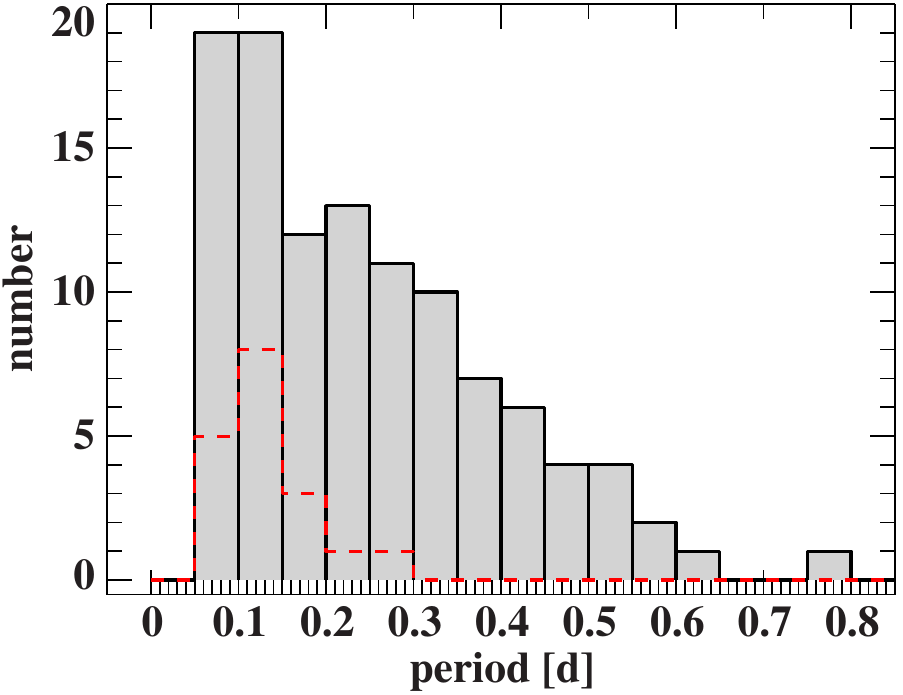}
	\caption{Period distribution of the OGLE HW Vir candidates in comparison to the known HW Vir systems. The OGLE systems are shown in grey and the known systems by the red dashed line.}
\end{figure}

It is not surprising that the number of systems at larger periods is smaller as the eclipse probability decreases with increasing separation ($P$(eclipsing = $(R_1+R_2)$/a). For the calculation of the eclipse probability we used the masses and radii of the sdB and M dwarf companion of the prototype HW Vir. The eclipse probability was then used to correct the period distribution of the OGLE targets, to derive the unbiased period distribution of sdB systems with cool companions. This is shown in Fig. 5. As not all systems have been confirmed yet, this figure is still preliminary. The period distribution shows a broad peak from 0.05 to 0.5 d with the maximum around 0.2 to 0.35 d. This means that the number of systems at longer periods is much larger than previously known.

\begin{figure}
	\label{period_corrected}
	\includegraphics[width=\linewidth]{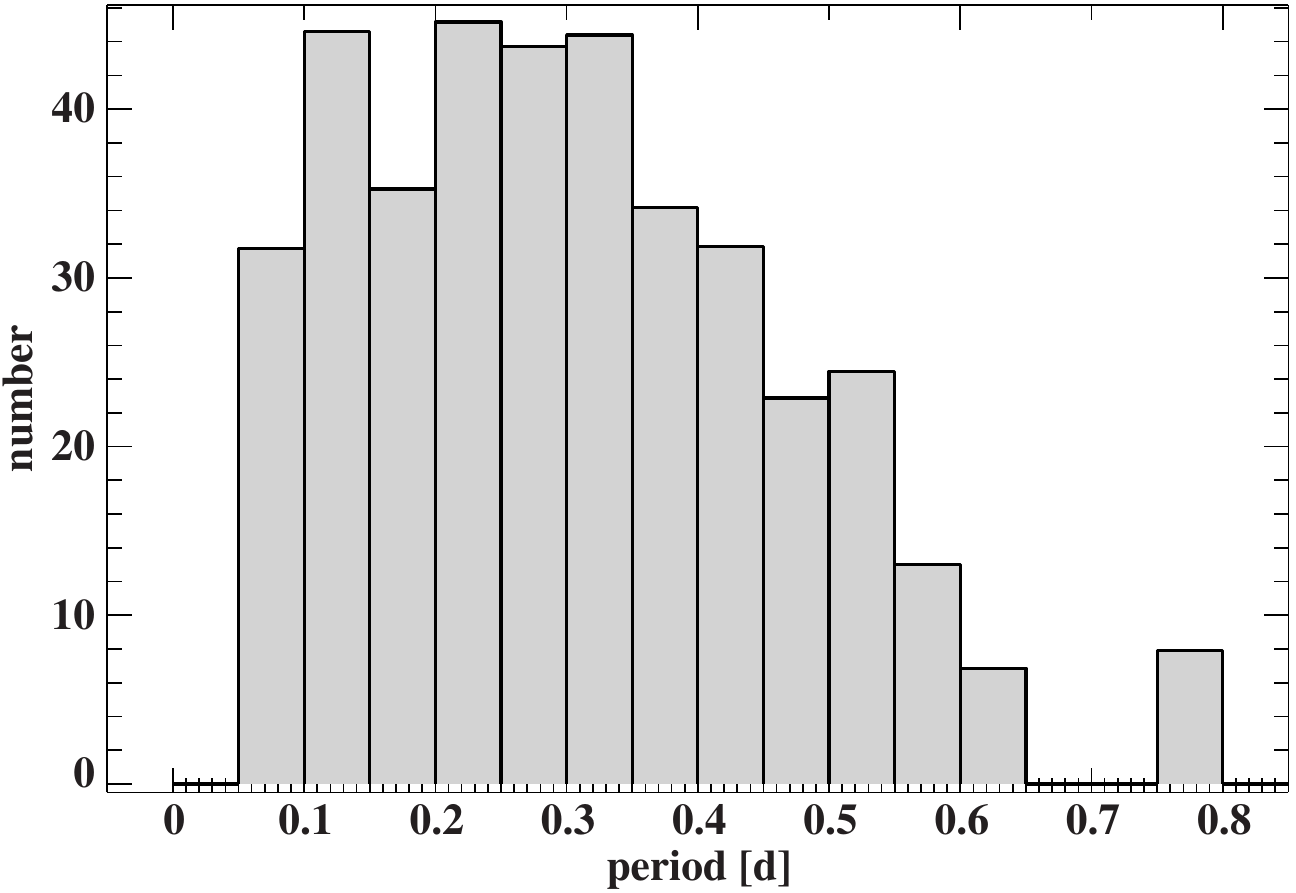}
	\caption{Preliminary period distribution of the OGLE HW Vir candidates corrected for the eclipse probability.}
\end{figure}

We used the spectral data of the observed OGLE targets to constrain the semi-amplitude of the radial velocity curve. This was combined with an analysis of the OGLE lightcurves to constrain the inclination angle. By assuming a typical mass of an sdB of 0.47 \Msun\, we were able to derive a preliminary period-companion mass diagram of the OGLE targets. This diagram is shown in Fig. 6. Similar to the period-companion mass diagram of the known HW Vir systems there is no correlation between companion mass and period. However, at shorter periods below 0.1 d the number of sub-stellar candidates is substantial and about equal to the number of M dwarf companions.

The spectra of the OGLE targets were also used to derive atmospheric parameters. These parameters were compared to the parameters of the known HW Vir systems (see Fig. 7). The diagram shows that the distribution of the OGLE HW Virs looks different in the $T_{\rm eff}-\log{g}$ diagram. The stars are distributed all over the extreme horizontal branch. This could mean that some of the OGLE HW Virs were formed by higher mass progenitors, while the known HW Virs, which are found in a very distinct area of the $T_{\rm eff}-\log{g}$ diagram, might have had similar progenitors. However, no targets at the hotter end of the EHB have been found yet.

\begin{figure}
	\label{logP-ogle}
	\includegraphics[angle=-90,width=\linewidth]{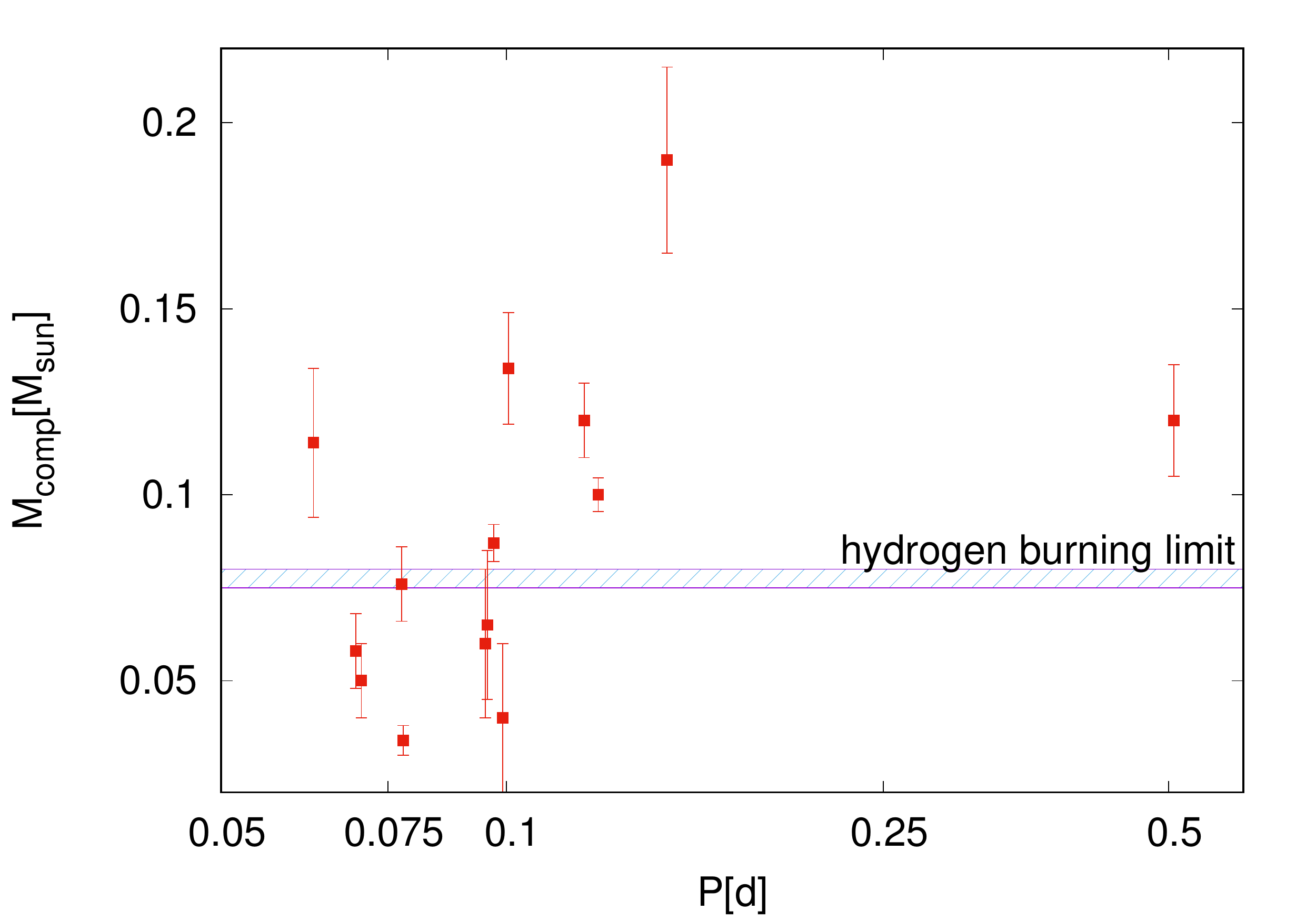}
	\caption{Preliminary period-companion mass diagram of the OGLE HW Vir systems with radial velocity curves.}
\end{figure}

\begin{figure}
	\label{}
	\includegraphics[width=\linewidth]{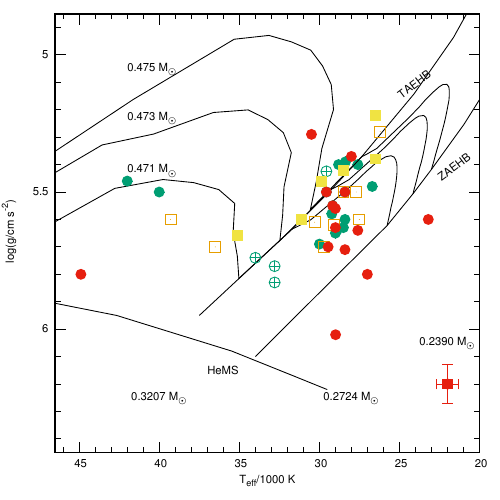}
	\caption{$T_{\rm eff}-\log{g}$ diagram of primaries of the reflection effect systems with known parameters. Open symbols represent pulsating systems. The green circles represent the known HW Vir systems, and the yellow squares represent the systems without eclipses. The red circles represent the analyzed HW Vir systems from the OGLE survey. The red square gives the typical statistical uncertainties.}
\end{figure}

\section{Conclusions}

We started the EREBOS project to study a large sample of homogeneously selected HW Vir systems from the OGLE survey. We plan a photometric and spectroscopic follow-up of as many targets as possible to determine the fundamental ($M$, $R$), atmospheric ($T_{\rm eff}$, $\log{g}$) and binary parameters ($a$, $P$). At the moment we already have RV data for 23 objects.

The data analysis turned out to be less straightforward than expected as we have some problems with the enormous crowding in the Galactic bulge field, which complicates photometry as well as spectroscopy. Moreover, the ESO-VLT/FORS2 spectrograph was not found to be as stable as expected, as it was not designed for the determination of radial velocities. This complicates the determination of radial velocity curves. 

The main goal is to investigate the systems at the short-period end and the longer-period range of the period distribution to answer our key questions:
Is there a well-defined minimum mass or a continuum ranging from the most
massive brown dwarfs down to hot Jupiter planets? What is the fraction of close sub-stellar companions to sdB
stars and how does it compare with the possible progenitor systems like main sequence stars with brown dwarf
or Jupiter-like companions? Moreover, we also need to get a better understanding of the common envelope phase. 
Another future goal is a physical model of the reflection effect, which we hope to achieve with this huge sample of reflection effect binaries. For this goal it is necessary to numerically model the radiation emitted from the heated hemisphere of the companion star rather than use a simple albedo approach as has been done in \citet{vuckovic:2016} for a large sample of systems.

\paragraph*{Acknowledgements}
V.S. is supported by the Deutsche Forschungsgemeinschaft (DFG) through grant GE2506/9-1. S.G. is supported by the Deutsche Forschungsgemeinschaft (DFG) through a Heisenberg fellowship (GE2506/8-1). Based on observations made with ESO Telescopes at the La Silla Paranal Observatory under programme ID 196.D-021 A-D, 095.D-0167, and 092.D-0040. Based on observations with SOAR/SOI under programme ID CN2016B-151 and CN2017A-100. This paper is based on observations obtained at the Southern Astrophysical Research (SOAR) telescope (NOAO Prop. ID: 2016A-0259 \& 2016B-0283; PI: B. Barlow), which is a joint project of the Ministério da Ciência, Tecnologia, e Inovação (MCTI) da República Federativa do Brasil, the U.S. National Optical Astronomy Observatory (NOAO), the University of North Carolina at Chapel Hill (UNC), and Michigan State University (MSU).
\bibliography{aabib}
\bibliographystyle{aa}
\appendix

\end{document}